\documentclass{article}
\usepackage{waspaa19,amsmath,amssymb,graphicx,url,times}
\usepackage{color}
\usepackage{enumitem}
\usepackage{multirow}
\usepackage{balance}
\usepackage{cite}

\def\thline{\noalign{\hrule height 1.0pt}}

\title{Cutting Music Source Separation Some Slakh:\\A Dataset to Study the Impact of Training Data Quality and Quantity}

\name{Ethan Manilow$^{1,2}$,
      Gordon Wichern$^{1}$,
      Prem Seetharaman$^{2}$,
      Jonathan Le Roux,$^{1}$\thanks{This work was performed while E.~Manilow was an intern at MERL.}}
\address{$^1$Mitsubishi Electric Research Laboratories (MERL), Cambridge, MA, USA\\ $^{2}$Interactive Audio Lab, Northwestern University, Evanston, IL, USA\\  
         {\small\texttt{\{ethanm, prem\}@u.northwestern.edu, \{wichern, leroux\}@merl.com}}
}

\begin{document}

\ninept
\maketitle

\begin{sloppy}

\begin{abstract}
Music source separation performance has greatly improved in recent years with the advent of approaches based on deep learning. Such methods typically require large amounts of labelled training data, which in the case of music consist of mixtures and corresponding instrument stems. However, stems are unavailable for most commercial music, and only limited datasets have so far been released to the public. It can thus be difficult to draw conclusions when comparing various source separation methods, as the difference in performance may stem as much from better data augmentation techniques or training tricks to alleviate the limited availability of training data, as from intrinsically better model architectures and objective functions.
In this paper, we present the synthesized Lakh dataset (Slakh) as a new tool for music source separation research. Slakh consists of high-quality renderings of instrumental mixtures and corresponding stems generated from the Lakh MIDI dataset (LMD) using professional-grade sample-based virtual instruments. A first version, Slakh2100, focuses on 2100 songs, resulting in 145 hours of mixtures. While not fully comparable because it is purely instrumental, this dataset contains an order of magnitude more data than MUSDB18, the {\it de facto} standard dataset in the field. 
We show that Slakh can be used to effectively augment existing datasets for musical instrument separation, while opening the door to a wide array of data-intensive music signal analysis tasks.
\end{abstract}

\section{Introduction}
\label{sec:intro}
Source separation performance has greatly improved in recent years thanks to the advent of approaches based on deep learning, starting with speech enhancement in non-stationary background noise~\cite{Erdogan2015}, then expanding to separation of voices from simultaneous overlapping speakers~\cite{Hershey2016} and separation of music mixtures~\cite{luo2017deep,jansson2017singing,Takahashi2018}. Such methods typically require large amounts of labelled training data, where the labels for a mixture are here the corresponding clean sources prior to mixing: in the case of speech enhancement and speech separation, where the target and interference are typically not correlated (or only loosely via the Lombard effect, for example), somewhat realistic data can be generated in large quantities by mixing randomly sampled examples of speech and noise or other speech. 
In the case of music, training data consists of mixtures and their corresponding instrument stems. However, stems are unavailable for most commercial music, and only limited datasets have thus far been released to the public. The largest publicly available dataset for music source separation, MUSDB18 \cite{musdb18}, is relatively small compared to datasets for other deep learning tasks: for example, MUSDB18 only contains 10 hours (from 150 songs) of mixture data, compared to 43 hours of mixtures in WSJ0-2mix, the most commonly used dataset for speech separation \cite{Hershey2016}.

Comparisons of methods that used data augmentation \cite{schluter2015exploring, uhlich2017improving} or external data \cite{pretet2019singing,jansson2017singing} show that models trained on more data typically perform higher on objective quality measures, such as source-to-distortion ratio (SDR). In fact, 4 of the top 5 performing algorithms at the SiSEC 2018 challenge \cite{stoter2018} used additional training data. This correlation between the amount and variety of training data and performance underlines the need for more training data. The impact of additional data on performance also underlines a fundamental problem regarding comparison across algorithms: is better performance due to the model, which is usually the emphasized novelty, or to better data augmentation and training tricks to alleviate data scarcity?

Furthermore, music features a large sonic variability, with a wider range of sounds, timbres, and pitches than there is in speech, for example. Existing datasets only partially address this problem, using terse categorization schemes for the sources in the mixture. For example, both MIR-1k \cite{chao-ling_hsu_improvement_2010} and iKala \cite{chan2015vocal} only provide sources for ``Vocals'' and ``Accompaniment''. While these categories have a historic impetus, they artificially circumscribe musical sources into two classes. MUSDB18 \cite{musdb18} and its older sibling DSD100 \cite{liutkus20172016} both use slightly more descriptive categories (``Vocals'', ``Bass'', ``Drums'', ``Other''), but this segmentation of musical instrument categories is still rudimentary. A large dataset with more granular categories could allow for source separation techniques that are more reflective of real-world musical conditions. 

Synthesizing a high-quality dataset could alleviate all of these problems, by allowing for fine-tuned control of the desired parameters. In fact, synthesized datasets have been shown to improve neural network performance on other music information retrieval tasks, such as drum transcription \cite{cartwright2018increasing},  understanding of compositional semantics and performance characteristics \cite{donahue2018nes}, synthesis of new sounds \cite{nsynth2017}, and frame-level recognition of instruments~\cite{hung2019multitask}. Previously, synthesized data was used in separating mixtures of stringed instruments~\cite{miron2017generating}, but we are unaware of any work that synthesizes audio for general music source separation.

In this paper, we present the synthesized Lakh dataset (Slakh), an open dataset consisting of high-quality renderings of instrumental mixtures and sources generated from the Lakh MIDI dataset (LMD) \cite{raffel2016learning}, and in particular a first public release focusing on 2100 songs, Slakh2100 (available at {\footnotesize \url{www.slakh.com}}). 
Using Slakh and a similar dataset generated using a simpler synthesizer called FluidSynth~\cite{fluidsynth}, referred to as Flakh, we analyze the impact of quantity and quality of training data on separation performance.

\section{Slakh dataset creation and analysis}\label{sec:page_size}
We propose to generate a large dataset of realistic instrumental mixtures and their corresponding stems by rendering MIDI files using high-quality  sample-based synthesis. Recordings in Slakh are generated using professional-grade virtual instruments used by countless musicians and composers. The dataset we release, Slakh2100, contains 2100 automatically mixed tracks and accompanying MIDI files separated into training (1500 tracks), validation (375 tracks), and testing (225 tracks) subsets, and totals 145 hours of mixtures. Additionally, the technique described here can lead to a virtually endless supply of high-quality mixtures and sources.

\subsection{Selection from the Lakh MIDI dataset}
\label{subsec:selection}
The Lakh MIDI dataset (LMD) \cite{raffel2016learning} is a collection of over 170,000 unique MIDI files scraped from the web. MIDI is a digital musical score that contains information for every note in a song, including information about what instruments should be used for each note. The MIDI file format segments different instruments into tracks similar to modern digital audio workstations (DAWs), which makes the notes for each instrument easy to isolate, and then synthesize.

Because the files of the LMD originate from thousands of unknown authors, there is no standardization between each file (including metadata, amount of instruments, instrument types or families). We therefore rely on the built-in ``instrument program numbers'' from the MIDI specification. These values prescribe instrument types and families to each track and are used to determine what virtual instruments to use when rendering audio for a given track.  Using the instrument program numbers, we select only files that contain instrument tracks for {\it at least} piano, bass, guitar, and drums, where each of these four instruments plays at least $50$ notes throughout the song. This subset contains $20,371$ MIDI files, but almost all of these selected files contain more than just those four instrument tracks. Once this subset is created, $2100$ files are randomly selected to be rendered . Note that the definition of this ``band'' (piano, bass, guitar, drums), and the minimum number of notes played by each instrument is easily altered to create mixtures with different characteristics. The number of mixtures that contain an instrument type by category is shown in Figure \ref{fig:mix_clas}.

\begin{figure}[t]
\includegraphics[width=0.99\columnwidth]{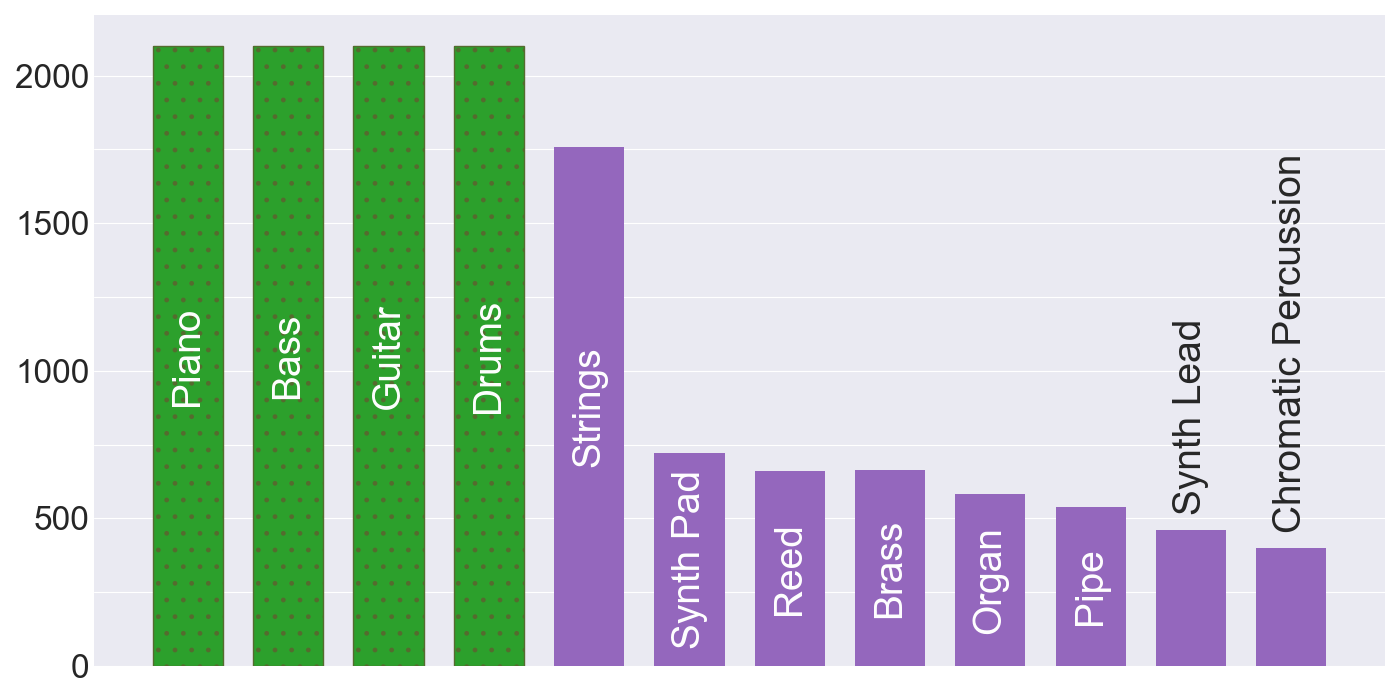}\vspace{-0.3cm}
\caption{Number of mixtures in Slakh2100 that contain at least one instrument from the following categories. Every mixture has {\it piano, bass, guitar,} and {\it drums} (the four leftmost bars, shown in green.)} \vspace{-0.3cm}
\label{fig:mix_clas}
\end{figure}

\subsection{Rendering and mixing}
\label{subsec:mixing}

Once the MIDI files are selected, we split each file up by creating a new MIDI file for each track in the original (or ``mixture'') MIDI file. The tracks are randomly assigned a patch that matches with its instrument program number. For instance, the MIDI instrument program numbers $0$ and $1$ correspond to ``Acoustic Grand Piano'' and ``Bright Acoustic Piano'', respectively. In this case, a patch is randomly selected from a list of twelve patches that fit the description of an acoustic piano. We use $187$ patches categorized into $34$ classes. These categories lead to $21,216$ possible instrument configurations for just the four instrument classes in our band; when we include other instruments, the number of possible configurations is immense. Many patches have effects built-in such as reverb, EQ, and compression. MIDI program numbers that are sparsely represented in LMD, like those under the ``Sound Effects'' header, are omitted. This selection and rendering process is repeated for every applicable track in the MIDI file. All tracks are rendered into separate monaural audio files at CD quality: 44.1kHz, 16-bit.

When making a recording in the studio, mixing is a time-consuming process done by ear by a recording engineer. Because we have neither the time nor budget to hire a professional to mix all 2100 songs, and because we want to use a method that could later scale up even further, we mix using an automatic procedure. We first normalize each track to be equal in terms of integrated loudness as calculated by the algorithm defined in the ITU-R BS.1770-4 specification \cite{BS1770}. Each track is then summed together instantaneously to make a mixture, and a uniform gain is applied to the mixture and each track to ensure that there is no clipping. This method for automatic mixing has been shown to be preferable to listeners in a subjective evaluation of automatic mixing techniques \cite{wichern2015comparison}. Creating stereo mixtures would require a suitable auto-mixing system to determine pan and delay for each track, which to the best of our knowledge does not yet exist.  
We note that the isolated Slakh instrument data is available for remixing.

\subsection{Flakh}

To study the impact of the quality of synthesis engines, the same 2100 MIDI files selected for Slakh2100 are also rendered with FluidSynth \cite{fluidsynth} using the `TimGM6mb.sf2' sound font, the default in pretty\_midi \cite{raffel2014intuitive}. We refer to the resulting dataset as Flakh. As before, we similarly split the MIDI into individual tracks and render these individually to make stems. But low-amplitude noise added by FluidSynth (most likely for dithering) can be boosted to unnaturally loud levels when normalizing and mixing Flakh as we did in Section \ref{subsec:mixing}. This makes it impossible to mix Flakh the same way as Slakh. Instead FluidSynth renders the whole, unsplit MIDI ``mixture'' as the resultant audio mixture. The rendering in Flakh is arguably much simpler than that used in Slakh.

\subsection{Details and analysis}

The resulting Slakh2100 dataset is larger than all of the other existing, open multi-track music source separation datasets combined, both when measuring by number of mixtures and amount of time. The number of tracks in Slakh2100 ranges from 4 to 48, with a median of 9, whereas MedleyDB is the only other dataset with a variable amount of tracks, ranging from 1 to 26 with a median of 5. MedleyDB is the only dataset in Table \ref{table:comparison} that is not explicitly for source separation, but part of it has been repackaged in MUSDB18 for this purpose. The LMD and its subset described in Section \ref{subsec:selection} both contain a nearly bottomless well of MIDI data to potentially generate more multi-track data. See Table \ref{table:comparison} for a full comparison.

\begin{table}[t]
\vspace{-0.24cm}
\caption{Comparison of open multi-track music datasets with the proposed Slakh2100 dataset. Size is measured in hours of mixture data. Data for the original Lakh MIDI dataset are also displayed to show the potential for further expansion of the Slakh dataset. $^\dagger$MUSDB18 combines songs from DSD100 and MedleyDB.}\label{table:comparison}
\vspace{-0.1cm}
\begin{center}
{\setlength{\tabcolsep}{4pt}
\begin{tabular}{l|r|r|r|r}
Dataset                          & \# Songs  & Size (h) & \# Tracks & \# Instr. cat. \\ \hline
iKala         \cite{chan2015vocal}                   & 306        & 2    & 2 & 2\\ 
MIR-1k   \cite{chao-ling_hsu_improvement_2010}                        & 110        & 2.25  & 2 & 2 \\ 
DSD100$^\dagger$      \cite{liutkus20172016}                     & 100        & 7    & 4 & 4 \\
MedleyDB$^\dagger$  \cite{bittner2014medleydb}       & 122 & 7.2 & 1--26 & 82\\ 
MUSDB18$^\dagger$\cite{musdb18}                          & 150        & 10    & 4 & 4 \\ 
\textbf{Slakh2100}               & 2100       & 145  & 4--48 & 34 \\ \thline
LMD (subset)                     & 20,371      & 1793   & 4+ & 129 \\ 
All LMD     \cite{raffel2016learning}                     & 176,581    & 10,521 & 1+ & 129 \\ 
\end{tabular}
}
\end{center}\vspace{-0.6cm}
\end{table}

Figure \ref{fig:spectra} shows a comparison of the mean and standard deviation over tracks of the normalized log-magnitude mixture spectra from MUSDB18, Slakh2100, and Flakh2100.  The average spectrum for a track is computed using an STFT with frame size of 4096 samples and 50\% overlap, averaging over all frames, and normalizing each average spectrum to have unit energy, as in the analysis of~\cite{pestana2013spectral}.  From Fig.~\ref{fig:spectra}, we note that the general spectral shape of the three datasets is quite similar, however, the mid-range harmonic peaks of the Slakh and Flakh datasets are much larger than for MUSDB18, since we use consistent tuning for the synthesized tracks.  Also, the low-frequency bump (around 80-200 Hz) is largely missing for Slakh and Flakh and they have a steeper drop in high frequency energy (around 10 kHz) compared to MUSDB18.  We also note that MUSDB18 is distributed in a lossy format causing the lack of energy above 16kHz.

The largest distinction between Slakh and existing datasets for music source separation is the absence of sources containing isolated vocals. This is due to the fact that the MIDI standard does not support vocals. While composers certainly do add ``vocal tracks'' using the available MIDI instrument programs, attempting to automatically infer which track is a vocal melody is beyond the scope of this paper. Although it might be obliquely possible to use this dataset to improve vocal isolation performance in other datasets by relegating vocals to a ``residual'' class, this limitation makes direct comparisons with existing datasets incomplete.

\section{Experimental validation}
\subsection{Model and setup}
We here train a separate network to recover each target instrument.
Let $X \in \mathbb{C}^{F\times T}$ be the complex spectrogram of a mixture of $C$ sources $S_c \in \mathbb{C}^{F\times T}$ for $c=1,\dots,C$. For simplicity, we focus here on methods which attempt to estimate a real-valued mask  $\hat{M}_c \in \mathbb{R}^{F\times T}$ for a single target source $c$ by minimizing the truncated phase sensitive approximation (tPSA) objective \cite{Erdogan2015}:
\begin{align}
	\mathcal{L}_{\text{tPSA}} =  \Big\| \hat{M}_{c} \odot |X|  - \operatorname{T}_{0}^{|X|}\left(|S_c| \odot \cos(\angle S_c - \angle X)\right) \Big\|_1, \label{eq:L_MI_PSA}
\end{align}
where $\odot$ denotes element-wise product, $\angle S_c$ is the true phase of source c, $\angle X$ is the mixture phase, and $\operatorname{T}_{0}^{|X|}(x)= \min(\max(x,0),|X|)$ is a truncation function ensuring the target can be reached with a sigmoid activation function.

\begin{figure}[t]
\includegraphics[width=0.99\columnwidth]{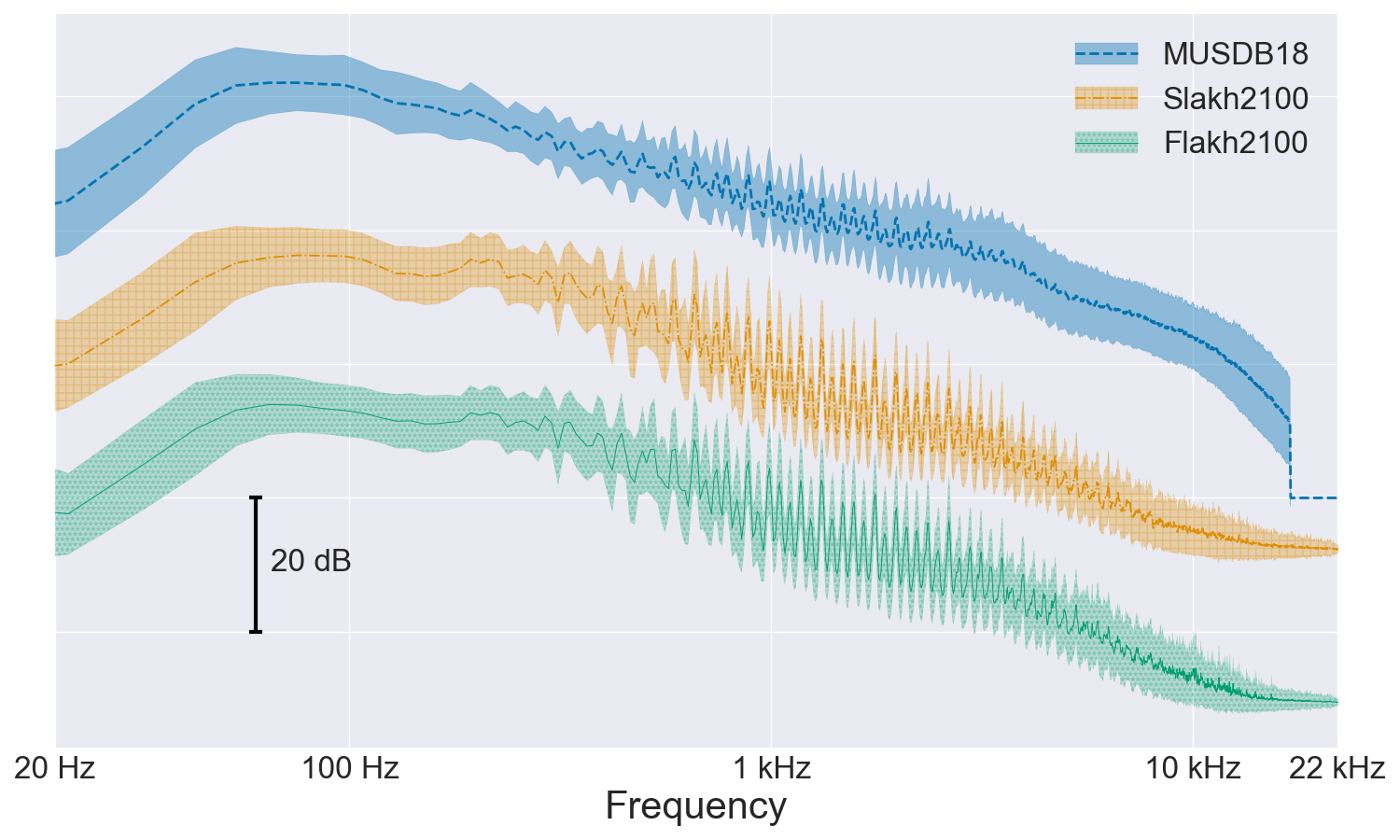}\vspace{-0.1cm}
\caption{Mean and standard deviation of all normalized log-magnitude mixture spectra in MUSDB18, Slakh2100, and Flakh2100.  For easy visual comparison, Slakh2100 and Flakh2100 are intentionally offset along the y-axis.}
\label{fig:spectra}\vspace{-0.2cm}
\end{figure}

For all experiments, we use a stack of 4 bidirectional long short-term memory (BLSTM) layers with 600 units in each direction followed by a dense output layer with sigmoid non-linearity, and log-magnitude spectrogram inputs. Other architectures such as chimera networks \cite{luo2017deep}, U-Nets \cite{jansson2017singing}, or DenseNets \cite{Takahashi2018} have been shown to outperform such a simple architecture, but our focus is here on investigating the influence of training data on performance in a controlled setting, and we shall thus leave the investigation of these models' performance using Slakh to future work. 

The networks are trained on 400-frame segments using the Adam algorithm, and dropout of 0.3 is applied to all BLSTM layers. All audio is downsampled to 16 kHz, and summed to mono for MUSDB18. The STFT window length is 32~ms and the hop size is 16~ms. The square root Hann window is employed as the analysis window, and the synthesis window is designed to achieve perfect reconstruction.  The loss function from Eq.~\eqref{eq:L_MI_PSA} is monitored on a validation set, and the learning rate is halved if the validation loss does not improve for $5$ consecutive epochs.  For MUSDB18, 86 tracks are used for training and 14 tracks for validation, while the standard 50 tracks are used for test.  Slakh2100 and Flakh2100 are split into 1500, 375, and 225 tracks in the training, validation, and test sets, respectively.
Performance is evaluated using the scale-invariant source-to-distortion ratio (SI-SDR) \cite{leroux2019sdr} between the reference source and the estimated source for each track in the given test set.

\subsection{Training sets}

We consider two ways to use the datasets to make training and validation sets for each target. The most straightforward way consists in considering the songs as they are: we refer to this case as ``coherent'', because all instrument sources in a mixture come from the same segment of a song and musically match with each other. In the ``coherent'' case, we build each sample by selecting a $10$-second clip starting at some random offset within a randomly selected song. The offset is constrained to be such that the target source had a presence above a certain RMS threshold in the clip, in this case -30 dB. The clips of the other sources for the corresponding time interval are then mixed (i.e., summed) together to create a submix of the interference in the mixture. 
Unless otherwise labeled, all datasets are mixed ``coherently'' using this procedure. 

An alternative consists in performing data augmentation on MUSDB18 by considering ``incoherent'' mixtures, where we no longer constrain all instruments to come from the same segment of the same song. This allows us to create large amounts of additional (unrealistic) data. 
Specifically, we build each sample by randomly selecting a 10-second (non-silent) clip for each MUSDB18 source (``Vocals'', ``Other'', ``Bass'' and ``Drums'')  at some random offset within some randomly selected song, where each source can be taken from a different song at a different offset. 
The  clips are mixed by scaling each source for equal loudness as in Sec.~\ref{subsec:mixing}. 

\subsection{MUSDB18 test set results}
\label{subsec:musdb_test}
In this section, we examine how using Slakh2100, Flakh2100, and a data augmented version of the MUSDB18 training set can lead to better results over the original MUSDB18 training set. We train models to separate ``Bass'' and ``Drums'',  because these are the only two sources shared between Slakh and MUSDB18.  Unless otherwise noted, all models use the MUSDB18 validation set to assess convergence, and we explore the following six training sets: (1) MUSDB18 without augmentation, (2) MUSDB18 augmented with Slakh2100 mixtures (MUSDB18 + Slakh), (3) MUSDB18 augmented with Flakh2100 mixtures (MUSDB18 + Flakh), (4) MUSDB18 augmented with remixed ``incoherent'' mixtures (MUSDB18 + MUSDB18-incoh.), (5) a reduced version of Slakh equal in size to the non-augmented MUSDB18, and using the Slakh2100 validation set (Slakh redux), and (6) a larger subset of Slakh2100 mixtures for both training and validation (Slakh).

\begin{table}[]
\footnotesize
\caption{Bass and drums separation performance in terms of SI-SDR [dB] averaged over the MUSDB18 test set for the unprocessed mixture, models trained on various datasets, and oracle methods.}\vspace{-.2cm}
\label{tab:results_musdb18}
\begin{center}
\begin{tabular}{l|c||c|c}
                             & Training     &       &\\
                             & data {[}h{]} & Bass  & Drums  \\ 
\hline \hline
Unprocessed mixture                 & -            & -6.0 & -3.8 \\ \hline 
MUSDB18                      & \phantom{4}5  & -0.5 & \phantom{-}2.2 \\ 
MUSDB18 + Slakh              & 48 & \phantom{-}1.3  & \phantom{-}3.6   \\ 
MUSDB18 + Flakh              & 48 & \phantom{-}0.0  & \phantom{-}3.1  \\ 
MUSDB18 + MUSDB18-incoh.     & 48 & \phantom{-}1.2  & \phantom{-}3.5  \\ 
Slakh redux                  & \phantom{4}5  & -2.0 & \phantom{-}0.6     \\ 
Slakh                        & 43 & -2.4 & \phantom{-}0.7     \\ 
\hline 
Oracle Methods:               &              &       &       \\
\quad IBM                          & -            & \phantom{-}5.9  & \phantom{-}8.4   \\
\quad IRM                          & -            & \phantom{-}5.7  & \phantom{-}8.1   \\
\quad Wiener-like                  & -            & \phantom{-}6.9  & \phantom{-}9.1   \\
\quad Truncated phase sensitive    & -            & \phantom{-}7.9  & 10.1    \\
\end{tabular}
\end{center}\vspace{-.6cm}
\end{table}

Table~\ref{tab:results_musdb18} compares the SI-SDR averaged over the MUSDB18 test set for each of the six training sets mentioned above, as well as the SI-SDR of the unprocessed mixture.  Additionally, Table~\ref{tab:results_musdb18} compares oracle mask performance (i.e., masks obtained from the ground truth signals) using the ideal ratio mask (IRM), ideal binary mask (IBM), Wiener filter-like mask, and the truncated phase sensitive filter, the oracle that most closely matches the phase sensitive objective function from Eq.~\eqref{eq:L_MI_PSA} \cite{Erdogan2015}.  We see that augmenting MUSDB18 leads to improvement over the non-augmented case, with Slakh2100 performing better than Flakh2100, showing that improving the realism of the synthesis system is important for source separation.  However, the gains between the MUSDB18 + Slakh and the simple augmentation in MUSDB18 + MUSDB18-incoh.\ are quite small (0.1 dB difference for both bass and drums).  While Slakh has the advantage of being musically consistent data, the augmented MUSDB18-incoh.\ data has the advantage of containing vocals and being a closer timbral match to the MUSDB18 test set.  Finally, we note that training on Slakh data alone does not lead to strong separation performance on the MUSDB18 test set, showing that this model overfits to its training data and generalizes poorly.

\subsection{Slakh2100 test set results}
Table~\ref{tab:results_slakh} compares performance on the 225 tracks in the Slakh2100 test set.  In addition to separating bass and drums, we also explore separation of guitar and piano. These instruments have previously been lumped into ``Other'' categories, but with Slakh we can easily benchmark existing source separation architectures on different non-vocal sources.  From Table~\ref{tab:results_slakh}, we see that as expected the Slakh and Slakh redux models perform better than the models trained on MUSDB18 for bass and drums separation.  We also notice that guitar and piano separation performance is quite low compared to bass and drums.  When listening to Slakh mixtures, the mid-frequency range occupied by guitar and piano has many overlapping instruments and it is difficult to clearly hear those sources.  Currently we only adjust levels when creating Slakh mixes as described in Section~\ref{subsec:mixing}, so introducing an automatic equalization system~\cite{hafezi2015autonomous} may help improve the quality of instruments such as guitar and piano in Slakh mixtures. We also note from Table~\ref{tab:results_slakh} that models trained without Slakh data (MUSDB18 and MUSDB18 + Flakh) perform poorly when separating bass. These results, along with similar results from Table \ref{tab:results_musdb18} showing poor generalization, indicate that models trained solely on one dataset have difficulties generalizing.

\begin{table}[]
\footnotesize
\caption{Separation performance in terms of SI-SDR [dB] averaged over the Slakh2100 test set for the unprocessed mixture, models trained on various datasets, and oracle methods.}\vspace{-.5cm}
\label{tab:results_slakh}
{\setlength{\tabcolsep}{3pt}
\begin{center}
\begin{tabular}{l|c||c|c|c|c}
                             & \!Training\! &       &       &        &       \\
                             & data {[}h{]} & Bass  & Drums & Guitar & Piano \\
\hline  \hline
Unprocessed mixture               & -            & -6.5 & -6.6 & -6.9  & -8.0 \\  \hline
MUSDB18                      & \phantom{4}5           & -3.2 & \phantom{-}3.2   & -      & -     \\
MUSDB18 + Slakh              & 48           & \phantom{-}3.3  & \phantom{-}9.0      & -      & -     \\
MUSDB18 + Flakh         & 48           & -10.2    & \phantom{-}4.1  & -      & -     \\
MUSDB18 + MUSDB18-incoh.     & 48           & \phantom{-}0.7    & \phantom{-}5.1  & -      & -     \\
Slakh redux                  & \phantom{4}5           & \phantom{-}3.3  & \phantom{-}9.4  & -3.0  & -3.1 \\
Slakh                        & 43           & \phantom{-}3.9  & \phantom{-}9.9  & -1.8  & -2.2 \\
\hline 
Oracle Methods:               &              &       &      && \\
\quad IBM                          & -            & 5.3   & 10.0   & 5.2 & 4.1 \\
\quad IRM                          & -            & 5.2   & 9.7    & 5.4 & 4.3 \\
\quad Wiener-like                  & -            & 6.3   & 10.7   & 6.4 & 5.3 \\
\quad Truncated phase sensitive    & -            & 7.3   & 11.7   & 7.3 & 6.4 \\
\end{tabular}
\end{center}
}\vspace{-.6cm}
\end{table}

\section{Conclusions and Future Work}
In this paper, we presented Slakh, a new open dataset for music source separation research consisting of high-quality renderings of instrumental mixtures and corresponding stems synthesized from MIDI using professional-grade sample-based virtual instruments. We showed how Slakh can be used to increase the amount of training data, which leads to improved performance, and also used to consider separation of new instrument categories that were not available in existing datasets. We note that with existing datasets, one needs to rely on unrealistic augmentations to achieve similar results to those achieved with Slakh. We also note that there is much room for growth with Slakh, such as synthesizing more mixtures or adding new instruments and effects, while the augmentation of existing datasets saturates comparatively quickly. Future work includes improving the quality and realism of Slakh mixes, for example via better equalization, mixing in isolated vocals from other datasets, and adding small deviations to note tuning. Future uses for Slakh include benchmarking stronger source separation models, as well as other data-intensive tasks like music transcription, instrument identification, and remixing for multi-channel separation.

\vfill\pagebreak

\bibliographystyle{IEEEtran}
\bibliography{refs19}

\begin{thebibliography}{10}
\providecommand{\url}[1]{#1}
\def\UrlFont{\rmfamily}
\providecommand{\newblock}{\relax}
\providecommand{\bibinfo}[2]{#2}
\providecommand\BIBentrySTDinterwordspacing{\spaceskip=0pt\relax}
\providecommand\BIBentryALTinterwordstretchfactor{4}
\providecommand\BIBentryALTinterwordspacing{\spaceskip=\fontdimen2\font plus
\BIBentryALTinterwordstretchfactor\fontdimen3\font minus
  \fontdimen4\font\relax}
\providecommand\BIBforeignlanguage[2]{{%
\expandafter\ifx\csname l@#1\endcsname\relax
\typeout{** WARNING: IEEEtran.bst: No hyphenation pattern has been}%
\typeout{** loaded for the language `#1'. Using the pattern for}%
\typeout{** the default language instead.}%
\else
\language=\csname l@#1\endcsname
\fi
#2}}

\bibitem{Erdogan2015}
H.~Erdogan, J.~R. Hershey, S.~Watanabe, and J.~{Le Roux}, ``Phase-sensitive and
  recognition-boosted speech separation using deep recurrent neural networks,''
  in \emph{Proc. IEEE International Conference on Acoustics, Speech and Signal
  Processing (ICASSP)}, Apr. 2015, pp. 708--712.

\bibitem{Hershey2016}
J.~R. Hershey, Z.~Chen, and J.~Le~Roux, ``Deep clustering: Discriminative
  embeddings for segmentation and separation,'' in \emph{Proc. IEEE
  International Conference on Acoustics, Speech and Signal Processing
  (ICASSP)}, Mar. 2016, pp. 31--35.

\bibitem{luo2017deep}
Y.~Luo, Z.~Chen, J.~R. Hershey, J.~Le~Roux, and N.~Mesgarani, ``Deep clustering
  and conventional networks for music separation: Stronger together,'' in
  \emph{Proc. IEEE International Conference on Acoustics, Speech and Signal
  Processing (ICASSP)}, Mar. 2017, pp. 61--65.

\bibitem{jansson2017singing}
A.~Jansson, E.~Humphrey, N.~Montecchio, R.~Bittner, A.~Kumar, and T.~Weyde,
  ``Singing voice separation with deep {U}-net convolutional networks,'' in
  \emph{Proc. International Society for Music Information Retrieval Conference
  (ISMIR)}, Oct. 2017.

\bibitem{Takahashi2018}
N.~Takahashi, N.~Goswami, and Y.~Mitsufuji, ``{MMDenseLSTM}: An efficient
  combination of convolutional and recurrent neural networks for audio source
  separation,'' in \emph{Proc. International Workshop on Acoustic Signal
  Enhancement (IWAENC)}, Sep. 2018.

\bibitem{musdb18}
\BIBentryALTinterwordspacing
Z.~Rafii, A.~Liutkus, F.-R. St{\"o}ter, S.~I. Mimilakis, and R.~Bittner, ``The
  {MUSDB18} corpus for music separation,'' Dec. 2017. [Online]. Available:
  \url{https://doi.org/10.5281/zenodo.1117372}
\BIBentrySTDinterwordspacing

\bibitem{schluter2015exploring}
J.~Schl{\"u}ter and T.~Grill, ``Exploring data augmentation for improved
  singing voice detection with neural networks.'' in \emph{Proc. International
  Society for Music Information Retrieval Conference (ISMIR)}, Oct. 2015, pp.
  121--126.

\bibitem{uhlich2017improving}
S.~Uhlich, M.~Porcu, F.~Giron, M.~Enenkl, T.~Kemp, N.~Takahashi, and
  Y.~Mitsufuji, ``Improving music source separation based on deep neural
  networks through data augmentation and network blending,'' in \emph{Proc.
  IEEE International Conference on Acoustics, Speech and Signal Processing
  (ICASSP)}, Mar. 2017, pp. 261--265.

\bibitem{pretet2019singing}
L.~Pr{\'e}tet, R.~Hennequin, J.~Royo-Letelier, and A.~Vaglio, ``Singing voice
  separation: A study on training data,'' in \emph{Proc. IEEE International
  Conference on Acoustics, Speech and Signal Processing (ICASSP)}, May 2019,
  pp. 506--510.

\bibitem{stoter2018}
F.-R. St{\"o}ter, A.~Liutkus, and N.~Ito, ``The 2018 signal separation
  evaluation campaign,'' in \emph{Proc. International Conference on Latent
  Variable Analysis and Signal Separation (LVA)}, 2018, pp. 293--305.

\bibitem{chao-ling_hsu_improvement_2010}
C.-L. Hsu and J.-S. Jang, ``On the improvement of singing voice separation for
  monaural recordings using the {MIR-1K} dataset,'' \emph{IEEE Transactions on
  Audio, Speech, and Language Processing}, vol.~18, no.~2, pp. 310--319, Feb.
  2010.

\bibitem{chan2015vocal}
T.-S. Chan, T.-C. Yeh, Z.-C. Fan, H.-W. Chen, L.~Su, Y.-H. Yang, and R.~Jang,
  ``Vocal activity informed singing voice separation with the {iKala}
  dataset,'' in \emph{Proc. IEEE International Conference on Acoustics, Speech
  and Signal Processing (ICASSP)}, Apr. 2015, pp. 718--722.

\bibitem{liutkus20172016}
A.~Liutkus, F.-R. St{\"o}ter, Z.~Rafii, D.~Kitamura, B.~Rivet, N.~Ito, N.~Ono,
  and J.~Fontecave, ``The 2016 signal separation evaluation campaign,'' in
  \emph{Proc. International Conference on Latent Variable Analysis and Signal
  Separation (LVA)}, Feb. 2017, pp. 323--332.

\bibitem{cartwright2018increasing}
M.~Cartwright and J.~P. Bello, ``Increasing drum transcription vocabulary using
  data synthesis,'' in \emph{Proc. International Conference on Digital Audio
  Effects (DAFx)}, Sep. 2018, pp. 72--79.

\bibitem{donahue2018nes}
C.~Donahue, H.~H. Mao, and J.~McAuley, ``The {NES} music database: A
  multi-instrumental dataset with expressive performance attributes,''
  \emph{arXiv preprint arXiv:1806.04278}, 2018.

\bibitem{nsynth2017}
J.~Engel, C.~Resnick, A.~Roberts, S.~Dieleman, D.~Eck, K.~Simonyan, and
  M.~Norouzi, ``Neural audio synthesis of musical notes with {WaveNet}
  autoencoders,'' \emph{arXiv preprint arXiv:1704.01279}, 2017.

\bibitem{hung2019multitask}
Y.-N. Hung, Y.-A. Chen, and Y.-H. Yang, ``Multitask learning for frame-level
  instrument recognition,'' in \emph{Proc. IEEE International Conference on
  Acoustics, Speech and Signal Processing (ICASSP)}, May 2019, pp. 381--385.

\bibitem{miron2017generating}
M.~Miron, J.~Janer~Mestres, and E.~G{\'o}mez~Guti{\'e}rrez, ``Generating data
  to train convolutional neural networks for classical music source
  separation,'' in \emph{Proc. Sound and Music Computing Conference (SMC)},
  Jul. 2017.

\bibitem{raffel2016learning}
C.~Raffel, ``Learning-based methods for comparing sequences, with applications
  to audio-to-midi alignment and matching,'' Ph.D. dissertation, Columbia
  University, 2016.

\bibitem{fluidsynth}
``{FluidSynth},'' \url{http://www.fluidsynth.org/}, {Online: Accessed April 30,
  2019}.

\bibitem{BS1770}
{Recommendation ITU-R BS.1770-4}, ``Algorithms to measure audio programme
  loudness and true-peak audio level,'' 2017.

\bibitem{wichern2015comparison}
G.~Wichern, A.~Wishnick, A.~Lukin, and H.~Robertson, ``Comparison of loudness
  features for automatic level adjustment in mixing,'' in \emph{Proc. Audio
  Engineering Society Convention}, Oct. 2015.

\bibitem{raffel2014intuitive}
C.~Raffel and D.~P.~W. Ellis, ``Intuitive analysis, creation and manipulation
  of midi data with pretty\_midi,'' in \emph{Proc. International Society for
  Music Information Retrieval Conference (ISMIR) Late Breaking and Demo
  Papers}, Oct. 2014, pp. 84--93.

\bibitem{bittner2014medleydb}
R.~M. Bittner, J.~Salamon, M.~Tierney, M.~Mauch, C.~Cannam, and J.~P. Bello,
  ``{MedleyDB}: A multitrack dataset for annotation-intensive {MIR} research,''
  in \emph{Proc. International Society for Music Information Retrieval
  Conference (ISMIR)}, Oct. 2014, pp. 155--160.

\bibitem{pestana2013spectral}
P.~D. Pestana, Z.~Ma, J.~D. Reiss, A.~Barbosa, and D.~A. Black, ``Spectral
  characteristics of popular commercial recordings 1950-2010,'' in \emph{Proc.
  Audio Engineering Society Convention}, Oct. 2013.

\bibitem{leroux2019sdr}
J.~Le~Roux, S.~Wisdom, H.~Erdogan, and J.~R. Hershey, ``{SDR}--half-baked or
  well done?'' in \emph{Proc. IEEE International Conference on Acoustics,
  Speech and Signal Processing (ICASSP)}, May 2019, pp. 626--630.

\bibitem{hafezi2015autonomous}
S.~Hafezi and J.~D. Reiss, ``Autonomous multitrack equalization based on
  masking reduction,'' \emph{Journal of the Audio Engineering Society},
  vol.~63, no.~5, pp. 312--323, 2015.

\end{thebibliography}

\end{sloppy}
\end{document}